\documentclass{article}
\usepackage{spconf,amsmath,amssymb,graphicx,hyperref}
\usepackage{booktabs} 
\usepackage{graphicx}
\usepackage[caption=false,font=footnotesize]{subfig} 
\usepackage[table]{xcolor}
\usepackage{multirow}
\usepackage{cite}
\usepackage{enumitem}

\title{Task Vector in TTS: Toward Emotionally Expressive Dialectal Speech Synthesis}
\name{\parbox{\linewidth}{\centering
Pengchao Feng$^{1,2*}$, 
Yao Xiao$^{1*}$, 
Ziyang Ma$^{1}$, 
Zhikang Niu$^{1,2}$, 
Shuai Fan$^{1}$, \\
Yao Li$^{3}$, 
Sheng Wang$^{1,3}$,  
Xie Chen$^{1,2\dagger}$
}
\thanks{$^*$ Equal contribution. $\dagger$ Corresponding Author.}
\thanks{The code and demo is available at \url{https://the-bird-f.github.io/Expressive-Vectors}.}
}

\address{
$^1$ School of Computer Science, Shanghai Jiao Tong University, China \\
$^2$ Shanghai Innovation Institute 
$^3$ Shanghai Aviation Electric Co., Ltd
}
\begin{document}
\ninept
\maketitle

\begin{abstract}
Recent advances in text-to-speech (TTS) have yielded remarkable improvements in naturalness and intelligibility. Building on these achievements, research has increasingly shifted toward enhancing the expressiveness of generated speech, such as dialectal and emotional TTS.
However, cross-style synthesis combining both dialect and emotion remains challenging and largely unexplored, mainly due to the scarcity of dialectal data with emotional labels. 
To address this, we propose \textbf{Hierarchical Expressive Vector (HE-Vector)}, a two-stage method for Emotional Dialectal TTS. 
In the first stage, we construct different task vectors to model dialectal and emotional styles independently, and then enhance single-style synthesis by adjusting their weights, a method we refer to as Expressive Vector (E-Vector).
For the second stage, we hierarchically integrate these vectors to achieve controllable emotionally expressive dialect synthesis without requiring jointly labeled data, corresponding to Hierarchical Expressive Vector (HE-Vector). 
Experimental results demonstrate that HE-Vectors achieve superior performance in dialect synthesis, and promising results in synthesizing emotionally expressive dialectal speech in a zero-shot setting. \end{abstract}
\section{Introduction}
\label{sec:intro}
In recent years, text-to-speech (TTS) technology has made remarkable progress, driven in large part by the availability of large TTS systems and scalable training datasets. Both autoregressive (AR) models \cite{wang2023neural, kim2024clam, peng2024voicecraft, anastassiou2024seed} and non-autoregressive (NAR) models \cite{chen2024f5, zhou2025indextts2, ju2024naturalspeech, du2024cosyvoice, du2024cosyvoice2,wang2025spark, zhang2025minimax} now achieve human-level speech quality and impressive zero-shot capabilities on unseen speakers. Building on these advances, there has been growing interest in enhancing the expressiveness of generated speech, with approaches falling into two categories: indirectly through the manipulation of objective acoustic features (e.g., latency, pitch, intensity) \cite{chen2025drawspeech}, or directly through the modeling of subjective expressive styles (e.g., dialect, emotion, speaking style) \cite{yang2025emovoice}. While acoustic features are relatively easy to model, direct control of expressive styles is substantially more challenging because of the weak alignment between abstract styles and acoustic spectra and the scarcity of high-quality labeled data. The challenge becomes even greater when jointly controlling multiple styles, as the scarcity of dialectal data with emotional labels and the potential interference among dialect, emotion, and other expressive factors further complicate the task.

To address these limitations, we propose the \textbf{Hierarchical Expressive Vector (HE-Vector)}, a two-stage method for both single-style and multi-style expressive speech synthesis. 
Specifically, in the first stage, we introduce \textbf{E-Vector}, an expressive style vector built upon F5-TTS, to capture the expressiveness of dialects and emotions individually. E-Vectors are derived from Task Vectors \cite{ilharco2022editing}, which amplifies style-specific features, improves clarity, and reduces interference from prompt audio. This method does not require full fine-tuning, offering high training efficiency.
In the second stage, we propose the \textbf{Hierarchically Merging Strategy} for integrating dialect and emotion E-Vectors. The key to this design is modulating dialect and emotion at separate layers of the model. This maximizes the effectiveness of each style's representation and ensures that learning one style does not interfere with the other. Crucially, this strategy requires no datasets with joint dialect-emotion labels, making it ideal for low-resource and zero-shot cross-style synthesis.

In summary, our contributions are threefold:
\begin{itemize}
    \item We propose HE-Vector, a two-stage framework that enables joint control of dialect and emotion without requiring datasets annotated with both attributes, improving flexibility and data efficiency.
    \item We introduce E-Vector, which linearly scales task vectors to enhance the characteristics of individual dialects or emotions, enabling efficient and clear single-style synthesis from limited data.
    \item We develop the hierarchical integration strategy, which controls dialect and emotion at separate model layers, allowing them to be trained independently and maximizing the effectiveness of each modulator.
\end{itemize}

\begin{figure*}[htb]
\centering
\subfloat[E-Vector Enhanced F5-TTS]{%
  \includegraphics[height=6cm]{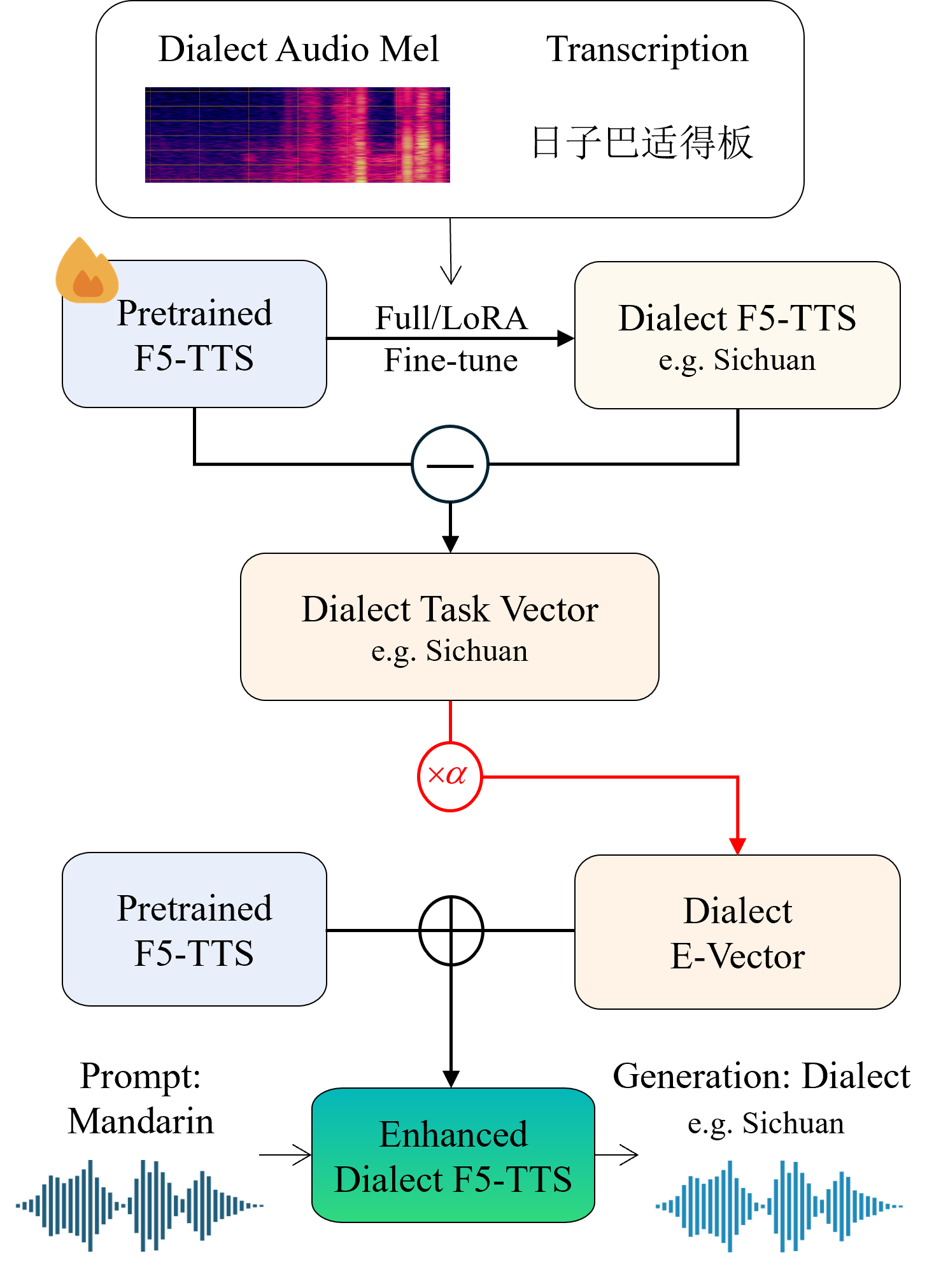}%
  \label{fig1}
}
\hspace{0.03\textwidth} 
\subfloat[Fully Merging Strategy]{%
  \includegraphics[height=6cm]{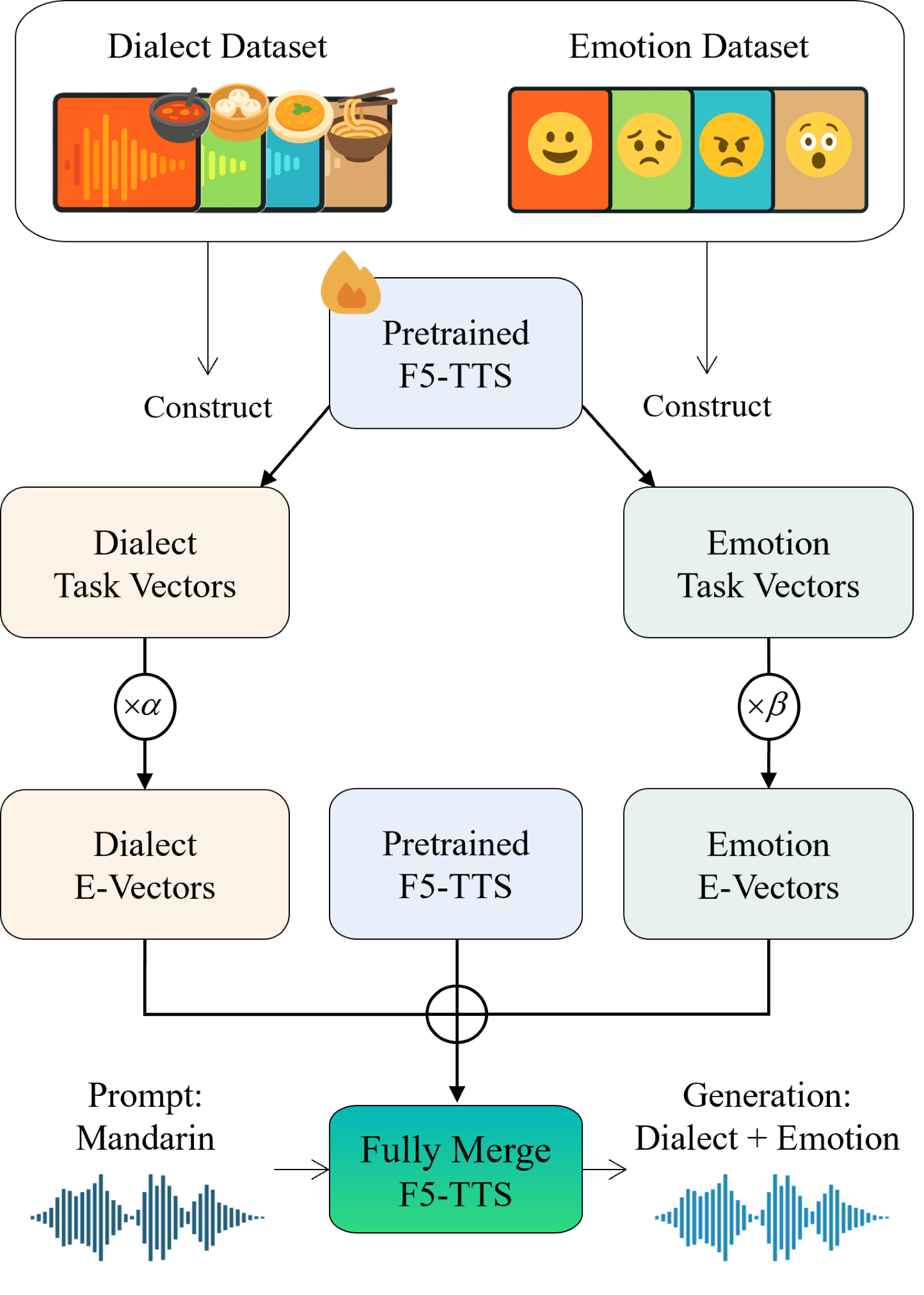}%
  \label{fig2}
}
\hspace{0.03\textwidth}
\subfloat[Hierarchically Merging Strategy]{%
  \includegraphics[height=6cm]{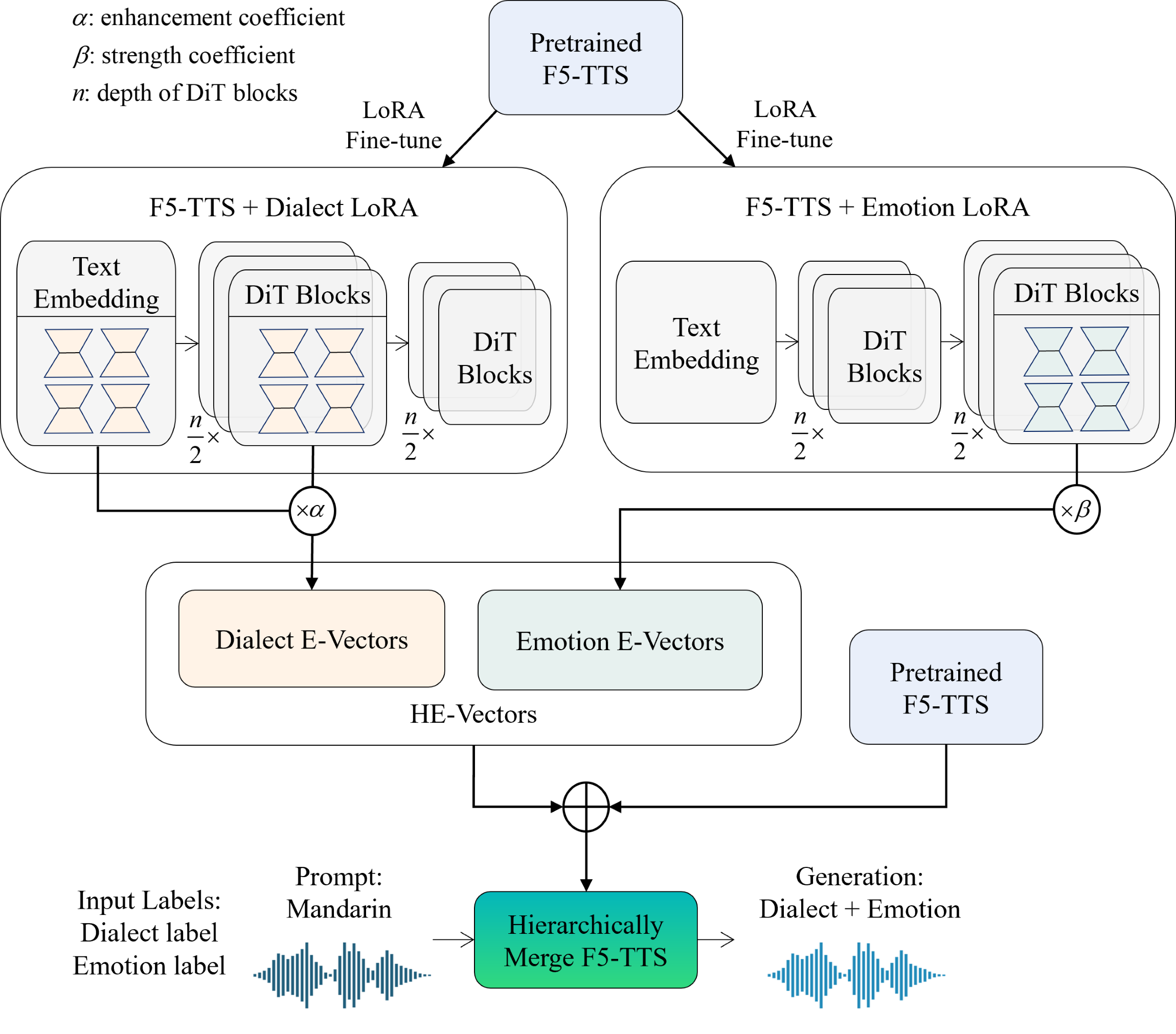}%
  \label{fig3}
}
\caption{Hierarchical Expressive Vector: 
(a) Construction of the E-Vector and enhancement of F5-TTS, 
(b) Fully merging strategy for dialect and emotion E-Vectors, 
(c) Hierarchically merging strategy for dialect and emotion E-Vectors}
\label{fig:res}
\end{figure*}

\section{Related Work}
\subsection{Dialect TTS and Emotion TTS}
Chinese dialects represent an important component of Chinese cultural heritage, and speech synthesis for dialects has received increasing attention. Zhang et al. \cite{zhang2022novel} proposed a Chinese dialect TTS frontend that converts Mandarin text into dialectal expressions, improving the intelligibility and naturalness of synthesized speech. Bailing TTS \cite{di2024bailing} was the first system to adopt a Mixture of Experts (MoE) architecture for zero-shot dialect synthesis. Beyond MoE-based approaches, the CosyVoice series \cite{du2024cosyvoice, du2024cosyvoice2} introduced an instruction-based framework that also supports high-quality zero-shot dialect synthesis, but these models face challenges in handling dialects with less clearly defined regional boundaries. 

Incorporating emotion into synthetic speech has long been a central focus in the field of TTS. Both coarse-grained models based on predefined emotion categories \cite{diatlova2023emospeech, guo2023emodiff} and fine-grained models leveraging natural language descriptions \cite{du2024cosyvoice, du2024cosyvoice2, yang2025emovoice} have demonstrated strong capabilities in generating emotionally expressive speech. However, due to the scarcity of dialectal speech corpora with reliable emotion annotations, the task of synthesizing emotional speech in dialects remains largely underexplored.

\subsection{Task Vector and Application}
Task Vector \cite{ilharco2022editing} is a modeling formulation of parameter variations that arise during fine-tuning, which can capture task-specific adaptation directions within the parameter space. This work first introduced the idea of using task vectors and the task algorithm to transfer deep neural networks to new tasks. Since then, task vectors have been widely applied in various domains, including capability editing in large language models \cite{huang2023chat}, low-resource speech recognition \cite{su2024task}, and unified modeling of music and speech synthesis \cite{ritter2025distilling}.  
Theoretical foundations of task vectors have also been strengthened. For example, Cheng \cite{cheng2025whoever} demonstrated the feasibility of linear-layer task vector composition.  
Motivated by these advances, we adopt task vectors to model subjective style capabilities, with dialects and emotions as representative cases. 

\section{Method}

\subsection{Expressive Vector (E-Vector)}
To efficiently capture the expressiveness of dialect or emotion, we construct E-Vector, which also forms the foundation for subsequent cross-style synthesis.

\subsubsection{Construct the E-Vector}
We construct the E-Vector upon F5-TTS. \textbf{F5-TTS} \cite{chen2024f5} is a zero-shot speech synthesis model with strong generalization ability, based on flow matching with a Diffusion Transformer (DiT). 

Taking dialect expression as an example, we first moderately fine-tune the pre-trained F5-TTS model on different dialect datasets. Then, as shown in Eq. (\ref{build}), we construct the dialectal task vectors by subtracting the parameters of the pre-trained model from those of the corresponding fine-tuned models.
\begin{equation}
\theta_{\text{pre}} \xrightarrow{\text{FT by i}} \theta_i, \  i \in \{\text{dialects}\}
\end{equation}
\begin{equation}
\tau_i = \theta_i - \theta_\text{pre},\quad \epsilon_i = \alpha \tau_i 
\label{build} 
\end{equation}

Here, $\theta \in \mathbb{R}^{n}$ ($n$ is the number of parameters) denotes the complete set of parameters of the F5-TTS model, with $\theta_{\text{pre}}$ representing the pretrained parameters and $\theta_{i}$ corresponding to the parameters fine-tuned for dialect $i$. The dialect task vector is denoted by $\tau_i \in \mathbb{R}^{n}$, and the corresponding dialect E-vector by $\epsilon_i \in \mathbb{R}^{n}$. $\alpha$ denotes the enhancement coefficient, which is determined based on validation results.

\subsubsection{Enhanced single-style synthesis via E-Vector}
Enhancement via E-Vector is based on the following two properties.

\textbf{Foundation.}
Lharco et al. \cite{ilharco2022editing} observed that the task vectors of a given pre-trained model and downstream tasks exhibit a consistent directional pattern within the parameter manifold. This directional consistency suggests that task vectors tend to converge toward a locally optimal solution. It serves as the foundation for constructing our dialect vector-enhanced model.

\textbf{Key factor.}
The parameter space of F5-TTS exhibits local insensitivity, as small perturbations (e.g., $\epsilon \sim \mathcal{N}(0,10^{-3})$) within a single DiT layer do not degrade perceptual quality. This robustness, similar to large language models \cite{huang2023chat}, enables F5-TTS to tolerate minor parameter perturbations without significant performance degradation, which is a key factor enabling our E-Vector enhanced model to achieve high-quality synthesis.

Specifically, as illustrated in Fig.~\ref{fig1}, by incorporating the dialect E-vector into the parameters of the pretrained model, we construct an enhanced dialect F5-TTS model that enables high-quality dialect synthesis. This approach explicitly models and reinforces the transferability of dialectal style, which can be seen as a type of Classifier-Free Guidance (CFG) \cite{ho2022classifier}. 

For attributes such as emotion, which exhibit continuous variation in contrast to categorical attributes like dialect, our approach enables controllable adjustment through the strength coefficient.
\begin{equation}
\epsilon_j = \beta \tau_j, \quad \beta \in [0, \beta_{\max}]
\end{equation}

Here, $\beta$ serves as the strength coefficient within a range, allowing explicit control over the intensity of a given emotion flexibly.  


\subsubsection{LoRA-based E-Vector}

Instead of applying full fine-tuning to the entire TTS model, we adopt LoRA \cite{hu2022lora} as a parameter-efficient alternative. Compared to full fine-tuning, LoRA not only reduces the number of trainable parameters but also allows multiple E-Vectors to coexist on a single backbone, supporting diverse styles without duplicating the model. 

To maximize their effectiveness, LoRA blocks are inserted into the modules that exhibit the largest parameter variations during full fine-tuning. Formally, let $W_\text{pre} \in \mathbb{R}^{d\times k}$ denote the frozen pre-trained weight of a module, which can be a linear, 1D convolutional, or embedding layer. For each dialect $i$, we associate an independent set of LoRA parameters $(A_i, B_i)$, where $A_i \in \mathbb{R}^{r\times k}$ and $B_i \in \mathbb{R}^{d\times r}$.

During training, the updated weights are computed as:
\begin{align}
W_i = W_\text{pre} + B_i A_i
\end{align}

At inference, we scale each dialect LoRA vector by the enhancement coefficient $\alpha$ to obtain the LoRA E-Vector:
\begin{align}
W_i = W_\text{pre} + \alpha^2 B_i A_i
\end{align}

\subsection{Hierarchical Expressive Vector (HE-Vector)}
In the previous section, we introduced the E-Vector, which models single-style expressiveness for dialect or emotion. However, generating speech with both styles requires an effective integration mechanism. Directly merging E-Vectors often leads to interference, so we propose the Hierarchical Expressive Vector (HE-Vector) framework, which introduces a hierarchical merging strategy, alongside a fully merged baseline for comparison.

\subsubsection{Fully Merging Strategy}
Following the Task Algorithm merging strategy \cite{ilharco2022editing}, as illustrated in Fig.~\ref{fig2}, the parameters of the dialect E-Vector and the emotion E-Vector are directly merged with the pretrained model parameters. While straightforward, this approach often leads to degraded controllability and audio quality due to style interference.

\subsubsection{Hierarchical Merging Strategy}
To mitigate these issues, we design a hierarchical merging strategy that assigns different control factors to different network layers, as shown in Fig.~\ref{fig3}. Specifically, a Dialect LoRA E-Vector is applied to the text embedding layer and the early half of the DiT blocks, where the model captures phonetic and pronunciation patterns most relevant to dialectal variation. An Emotion LoRA E-Vector is applied to the latter half of the DiT blocks, where control primarily shapes prosody, rhythm, and intonation.

At inference, these two LoRA E-Vectors are jointly applied to the pretrained backbone, each acting on its designated layers. This hierarchical composition allows dialect and emotion to be integrated without interference, ensuring that the two styles complement rather than override each other. Compared to fully merged approaches, this strategy achieves more stable cross-style control while maintaining audio quality.

\section{Experiments}

\subsection{Experiments Configuration}
\textbf{Datasets.} 
We used a dialect corpus in-house, covering 8 dialects with 10 hours of speech and transcripts per dialect, split into training/validation/test sets (8:1:1).  
The Emotion Speech Data \cite{zhou2022emotional} corpus was also adopted, and we partitioned it with the same 8:1:1 ratio.  
In addition, subsets of CV3-Eval \cite{gao2025differentiablerewardoptimizationllm} were used for evaluation. 
\begin{table}[ht]
\centering
\renewcommand{\arraystretch}{1.2}
\caption{Speech dataset used in our experiments.}
\resizebox{0.48\textwidth}{!}{%
\begin{tabular}{c|cc|cc}
\hline
\textbf{Corpus} & \textbf{Subset} & \textbf{Duration} & \textbf{Subset} & \textbf{Duration} \\
\hline
\multirow{4}{*}{\shortstack{Dialect Corpus \\ (in-house)}} 
& Tianjin & 10.00 h & Henan & 10.00 h \\
& Guangdong  & 10.00 h & Shaanxi & 10.00 h \\
& Shanghai    & 10.00 h & Hunan   & 10.00 h \\
& Sichuan    & 10.00 h &  Shandong  & 10.00 h \\
\hline
Emotion Speech & Happy & 5.38 h & Sad & 6.83 h \\
Dataset & Angry & 5.33 h & Surprise & 5.88 h \\
\hline
\end{tabular}
}
\label{tab:corpora}
\end{table}

\textbf{Evaluation Metrics.} 
Subjective metric is \textbf{Mean Opinion Score (MOS)} ratings for the overall naturalness (whether the speech matches the intended description and achieves good perceptual quality). Each dialect evaluation was conducted by more than five raters who are native to or highly familiar with the corresponding dialect. 
Objective metrics include (1) \textbf{Word Error Rate (WER)}, computed by transcribing synthesized speech with a Seed ASR \cite{bai2024seed} and aligning with the reference text; (2) \textbf{Speaker similarity (SIM-O)}, measured with the 3D-Speaker model \cite{chen20253d}.

\subsection{Dialect Synthesis}
\begin{table*}[ht]
\centering
\renewcommand{\arraystretch}{1.1}
\caption{Subjective evaluation of \textbf{Dialect Synthesis with Mandarin prompts} (mean $\pm$ std) across different dialects, with row-wise averages.}
\label{T1}
\resizebox{\textwidth}{!}{%
\begin{tabular}{c|cccccccc|c}
\toprule
\textbf{Method} & \textbf{Tianjin}  & \textbf{Guangdong} & \textbf{Shanghai} & \textbf{Sichuan} & \textbf{Henan}  &  \textbf{Shaanxi} & \textbf{Hunan}& \textbf{Shandong} & \textbf{Avg.} \\
\midrule
\rowcolor{gray!20} GT & 3.50 $\pm$ 1.27  & 4.03 $\pm$ 0.98 & 3.54 $\pm$ 1.03 & 3.70 $\pm$ 0.95 & 3.75 $\pm$ 0.82  & 3.74 $\pm$ 1.18 &  3.25 $\pm$ 0.86  & 3.99 $\pm$ 1.00 & 3.69 \\

CosyVoice2 & 2.70 $\pm$ 1.39 & \textbf{3.65 $\pm$ 1.12} & 3.03 $\pm$ 0.86 & 3.30 $\pm$ 0.98 & 2.11 $\pm$ 1.09  & 1.71 $\pm$ 0.98  & \textbf{ 2.92 $\pm$ 1.15} & 1.56 $\pm$ 0.96 & 2.62 \\

FT & 1.76 $\pm$ 1.14 & 1.31 $\pm$ 0.45 & 1.51 $\pm$ 0.71 & 1.96 $\pm$ 0.81 & 2.50 $\pm$ 0.97  & 1.99 $\pm$ 1.00 &  1.42 $\pm$ 0.60 & 2.34 $\pm$ 0.96  & 1.85 \\
FT-last & \textbf{3.16 $\pm$ 1.06 } & 3.53 $\pm$ 1.19  & 2.05 $\pm$ 0.93 & 2.97 $\pm$ 0.95 & 3.29 $\pm$ 0.74 & 1.88 $\pm$ 0.92 &  2.78 $\pm$ 0.65 & 3.11 $\pm$ 0.87 & 2.85 \\

E-Vector & 3.07 $\pm$ 1.02  & 2.99 $\pm$ 1.19  & \textbf{3.46 $\pm$ 0.92} & \textbf{3.51 $\pm$ 0.92} & \textbf{3.30 $\pm$ 0.78} & \textbf{3.44 $\pm$ 1.16} &  2.23 $\pm$ 0.89 & \textbf{3.49 $\pm$ 0.94} & \textbf{3.18} \\

LoRA E-Vector & 2.19 $\pm$ 1.10 & 1.54 $\pm$ 0.65  & 2.18 $\pm$ 0.89  & 2.54 $\pm$ 0.94 & 2.98 $\pm$ 0.79& 2.77 $\pm$ 1.22 &  1.52 $\pm$ 0.71 & 3.09 $\pm$ 0.93 & 2.35 \\
\bottomrule
\end{tabular}
}
\end{table*}

\begin{table*}[ht]
\centering
\renewcommand{\arraystretch}{1.1}
\caption{Subjective evaluation of \textbf{Emotional Dialect Synthesis} (mean $\pm$ std) across different dialects, with row-wise averages.}
\label{T3}
\resizebox{\textwidth}{!}{%
\begin{tabular}{c|cccccccc|c}
\toprule
\textbf{Method} & \textbf{Tianjin} & \textbf{Guangdong} & \textbf{Shanghai} & \textbf{Sichuan} & \textbf{Henan} & \textbf{Shaanxi} & \textbf{Hunan} & \textbf{Shandong} & \textbf{Avg.} \\
\midrule
CosyVoice2 & 1.74 $\pm$ 0.92 & \textbf{2.60 $\pm$ 1.11} & 2.13 $\pm$ 0.99 & 2.74 $\pm$ 1.06 & 1.59 $\pm$ 0.79 & 1.35 $\pm$ 0.60 & 1.65 $\pm$ 0.83 & 1.18 $\pm$ 0.33 & 1.87\\
Dual-stage & 2.31 $\pm$ 0.97 & 2.04 $\pm$ 0.94 & 2.61 $\pm$ 0.91 & 2.86 $\pm$ 0.95 & 2.83 $\pm$ 1.03 & \textbf{2.88 $\pm$ 0.98} & 2.28 $\pm$ 1.08 & 2.70 $\pm$ 1.02 & 2.56 \\
Fully E-Vector & \textbf{2.75 $\pm$ 1.07} & 2.09 $\pm$ 0.90 & 2.99 $\pm$ 0.91 & \textbf{2.97 $\pm$ 0.71} & 3.06 $\pm$ 0.78 & 2.63 $\pm$ 0.93 & 2.50 $\pm$ 0.79 & 3.08 $\pm$ 1.00 & 2.76 \\
HE-Vector & 2.68 $\pm$ 0.95 & 1.73 $\pm$ 0.80 & \textbf{3.07 $\pm$ 0.77} & 2.61 $\pm$ 0.80 & \textbf{3.28 $\pm$ 0.79} & 2.80 $\pm$ 0.90 & \textbf{3.22 $\pm$ 0.70} & \textbf{3.26 $\pm$ 0.71} & \textbf{2.83} \\
\bottomrule
\end{tabular}
}
\end{table*}

\begin{table}[ht]
\centering
\renewcommand{\arraystretch}{1.2}
\caption{Objective evaluation of \textbf{Dialect Synthesis with Mandarin prompts}. $^*$ASR evaluation covers only Guangdong, Shanghai, Sichuan, and Shaanxi dialects due to Seed-ASR constraints, with potential recognition errors.}
\label{T2}
\begin{tabular}{c|cc}
\toprule
\textbf{Method} & \textbf{Avg. WER($\%$)}$^*$ $\downarrow$ & \textbf{Avg. SIM-O} $\uparrow$ \\
\midrule
\rowcolor{gray!20} GT & 16.59 & - \\
CosyVoice2        & 14.49 &  0.63 \\
FT          & 9.04 &  0.72 \\
FT-Last        & 7.43 &  0.65 \\
E-Vector    & 15.41 &  0.65 \\
LoRA E-Vector  &18.58 &  0.70 \\
\bottomrule
\end{tabular}
\end{table}

The dialect synthesis task can be divided into two settings: (1) \textbf{Easy Task}, synthesizing dialectal speech from a dialectal prompt, and (2) \textbf{Hard Task}, synthesizing dialectal speech from a Mandarin prompt. Since each model can achieve comparable results to the ground truth in the former aspect, we are more focused on the latter. 

For comparison, we evaluate our method against several baselines: (1) \textbf{CosyVoice2} \cite{du2024cosyvoice2}: one of the few open-source models capable of zero-shot dialectal speech synthesis;
(2) \textbf{FT}: an F5-TTS model fine-tuned for 60k steps;
(3) \textbf{FT-last}: an over-fine-tuned F5-TTS model (trained until the validation loss plateaued, approximately 340k steps);
(4) \textbf{E-Vector}: our proposed method (F5-TTS fine-tuned for 60k steps, enhancement coefficient $\alpha = 3.0$, which was selected based on the subjective results obtained from the validation set);
(5) \textbf{LoRA E-Vector}: an alternative version that leverages LoRA-based modeling of the E-Vector (the enhancement coefficient $\alpha = 1.12$, the LoRA rank $r=8$, which provides a favorable trade-off between expressiveness and parameter efficiency.). Both this experiment and the Emotion TTS experiments are presented on our demo page.

As shown in Table \ref{T1}, the E-Vector Enhanced model achieves the highest average MOS, outperforming CosyVoice2, which was trained on thousands of hours of speech data.
This result highlights both the efficiency of E-Vector in leveraging limited data and the advantage of expert models over general-purpose models. 
Remarkably, it requires only one-fifth of the training steps of the over-fine-tuned F5-TTS model to achieve high-quality dialect synthesis, indicating that the method also accelerates convergence during fine-tuning. 


As shown in Table \ref{T2}, the objective evaluation should be interpreted in terms of relative magnitude, since the adopted evaluation tools introduce certain errors in dialectal speech recognition. The results indicate that our method achieves WER and speaker similarity comparable to other approaches (and even to the ground truth), which demonstrates that the E-Vector does not compromise either the correctness of synthesized speech or the preservation of speaker characteristics.


\subsection{Emotional Expressive Dialectal Speech Synthesis}

In this task, we focus on emotional dialectal speech synthesis, where a Mandarin reference audio is provided along with the target dialect and emotion labels as synthesis conditions. The generated speech samples, together with their corresponding style descriptions, are then evaluated via subjective listening tests.

The experimental comparison involves several representative systems, which provide a comprehensive basis for evaluating the effectiveness of our proposed framework:  
(1) \textbf{CosyVoice2}: one of the few open-source models capable of instruction-based multi-style speech synthesis;  
(2) \textbf{Dual-stage pipeline}: an engineering approach where dialect-enhanced F5-TTS and emotion-enhanced F5-TTS are sequentially combined to produce emotional dialectal speech;  
(3) \textbf{Fully E-Vector}: a method that integrates dialect and emotion E-vectors using a fully merging strategy;  
(4) \textbf{HE-Vector}: our proposed approach that integrates multiple task vectors using a hierarchical merging strategy to improve synthesis quality while maintaining controllability.  

As shown in Table \ref{T3}, the HE-Vector achieves the best overall quality, followed by the Fully E-Vector. This first demonstrates the feasibility of fully merging E-Vectors, serving as an empirical validation of the Task Algorithm. More importantly, it highlights the advantages of our hierarchical merging strategy, which not only mitigates the error accumulation caused by directly combining different E-Vectors but also reduces the parameter overhead, making it particularly well-suited for MoE models. At the same time, the results also reveal that existing models often fail when attempting to simultaneously control two or more expressive styles, highlighting the challenging nature of this research problem.


\section{Discussion}
\textbf{E-Vector for other TTS models.}
We also applied this approach to CosyVoice \cite{du2024cosyvoice}, but found degraded synthesis quality. This is mainly because expressive vector enhancement interferes with the coordination between its LLM-based text encoder component and the flow-matching acoustic model component.

\noindent\textbf{Strategy for construction and merging.}
Our analysis of parameter variations during expressive style transfer fine-tuning reveals that the shifts are not strictly linear, indicating a limitation of the E-Vector construction with linear scaling. Assigning different coefficients to DiT layers also brought no significant gain. Developing more effective strategies for constructing and merging E-Vectors remains an important direction for future work.

\section{Conclusion}
In this paper, we aim to tackle the novel problem of synthesizing emotionally expressive dialectal speech. We presented \textbf{Hierarchical Expressive Vector (HE-Vector)}, a two-stage approach for emotional dialectal TTS. By independently modeling dialectal and emotional styles as E-Vectors in the first stage and hierarchically integrating them in the second stage, HE-Vector enables controllable synthesis without requiring jointly labeled data. Experimental results validate the effectiveness of our method in dialect synthesis and demonstrate its potential for broader expressive style control. We believe this work represents an important step toward flexible and data-efficient expressive speech synthesis, paving the way for future research on multi-style speech generation.

\bibliographystyle{IEEEbib}
\bibliography{strings,refs}

\end{document}